\documentclass[journal]{IEEEtran}

\usepackage{amsmath,amssymb}
\usepackage{graphicx}
\usepackage{array}           
\usepackage{booktabs}        
\usepackage{textcomp}        
\usepackage{xcolor}
\usepackage{bm}
\usepackage{url}
\usepackage{listings}

\definecolor{lstkw}{HTML}{7A3EA8}    
\definecolor{lstfn}{HTML}{2F6F9F}    
\definecolor{lststr}{HTML}{2E7D32}   
\definecolor{lstnum}{HTML}{B5451F}   
\definecolor{lstcom}{HTML}{8A929B}   
\definecolor{lstgut}{HTML}{B3BCC4}   
\definecolor{lstframe}{HTML}{CFD6DC} 
\lstdefinestyle{pyfig}{
  language=Python,
  basicstyle=\ttfamily\scriptsize,
  keywordstyle=\color{lstkw}\bfseries,
  commentstyle=\color{lstcom}\itshape,
  stringstyle=\color{lststr},
  identifierstyle=\color{black},
  emph={range,scans,plot_2d,update_2d,awg_next_phase,pulser_update,%
        digitizer_get_curve,pulser_shift,awg_shift,pulser_pulse_reset,%
        awg_pulse_reset},
  emphstyle=\color{lstfn},
  numbers=left, numberstyle=\ttfamily\scriptsize\color{lstgut}, numbersep=8pt,
  frame=single, rulecolor=\color{lstframe}, framesep=4pt,
  showstringspaces=false, columns=fullflexible, keepspaces=true,
  xleftmargin=14pt, xrightmargin=8pt, aboveskip=2pt, belowskip=2pt,
}

\newcommand{\SI}[2]{#1\,\text{#2}}
\newcommand{\SIrange}[3]{#1--#2\,\text{#3}}
\providecommand{\ohm}{\ensuremath{\Omega}}
\providecommand{\micro}{\textmu}
\providecommand{\litre}{L}

\graphicspath{{figures/}}

\begin{document}

\title{Highly Efficient 3-in-1 FPGA-Based Unit Combining an ADC, DAC,
       and Multichannel Pulse Generator for Pulsed EPR Spectroscopy}

\author{N.~P.~Isaev,
        A.~R.~Melnikov,
        A.~V.~Vedkal,
        A.~S.~Ishchenko,
        A.~M.~Sokolov,
        E.~A.~Voronkov,
        and~S.~L.~Veber
\thanks{Manuscript received XX Month 2026; revised XX Month 2026; accepted
  XX Month 2026. \emph{(N. P. Isaev and A. R. Melnikov contributed equally to
  this work.)} \emph{(Corresponding authors: N. P. Isaev;
  A. R. Melnikov.)}}%
\thanks{This work was supported in part by the Russian Science Foundation
  under Grant 23-73-00042.}%
\thanks{This article has supplementary downloadable material provided by the
  authors: a complete Atomize script for the four-pulse DEER measurement used
  for the benchmarks reported here. It is available as an ancillary file
  accompanying this preprint.}%
\thanks{N. P. Isaev is with the Voevodsky Institute of Chemical Kinetics and
  Combustion, Siberian Branch, Russian Academy of Sciences, 630090
  Novosibirsk, Russia (e-mail: isaev@kinetics.nsc.ru).}%
\thanks{A. R. Melnikov and S. L. Veber are with the International Tomography
  Center, Siberian Branch, Russian Academy of Sciences, 630090 Novosibirsk,
  Russia, and also with the Synchrotron Radiation Facility ``SKIF,'' 630090
  Novosibirsk, Russia (e-mail: anatoly.melnikov@tomo.nsc.ru).}%
\thanks{A. V. Vedkal is with the Synchrotron Radiation Facility ``SKIF,''
  630090 Novosibirsk, Russia.}%
\thanks{A. S. Ishchenko is with the International Tomography Center, Siberian
  Branch, Russian Academy of Sciences, 630090 Novosibirsk, Russia, and also
  with Novosibirsk State University, 630090 Novosibirsk, Russia.}%
\thanks{A. M. Sokolov and E. A. Voronkov are with Instrumental Systems,
  117587 Moscow, Russia.}}

\bstctlcite{IEEEbstctl}

\maketitle

\begin{abstract}
A hardware and software system for pulsed electron paramagnetic resonance (EPR)
spectroscopy is presented, built around a single PCI~Express (PCIe) card that
combines a 16-bit digital-to-analog converter (DAC) with a \SI{10}{GS/s}
effective sample rate, a
\SI{2.5}{GS/s} 14-bit analog-to-digital converter (ADC), and an 11-channel pulse
generator on board. Custom firmware reloads the pulse-generator and DAC buffers
and streams the digitized data in parallel with a running experiment, so that no
time is lost on data input or output. The unit is integrated into the
open-source Atomize software and controlled at the level of pulse sequences
through a high-level Python application programming interface (API).
Benchmarking with a four-pulse double electron--electron resonance (DEER) sequence 
shows that the card reaches close to 100\% time efficiency for accumulation times 
of a few milliseconds per pulse sequence, even for demanding arbitrary waveform 
generator (AWG) protocols. Although demonstrated for a single sequence, the 
efficiency proved insensitive to the number of pulses, the number of measurement 
points, or the phase-cycling depth, so that comparable performance is expected 
for any protocol whose shot repetition time exceeds the ${\sim}$\SI{2}{ms} reprogramming time
of the card. The on-the-fly reprogramming, which makes every successive sequence fully
independent, also enables nonuniform sampling and acquisition. The system design, its firmware,
the Python API, and its performance in pulsed EPR applications are described.
\end{abstract}


\IEEEpeerreviewmaketitle


\raggedbottom

\section{Introduction}
Pulsed electron paramagnetic resonance (EPR) spectroscopy is a powerful tool
for studying the properties of paramagnetic centers, their nuclear environment, and nanometer distances
between them using pulsed dipolar spectroscopy (PDS)~\cite{pds_review_a,pds_review_b,pds_review_c}. PDS is
primarily used in structural biology to measure distances between spin-labeled
sites on proteins~\cite{pds_biology_a,pds_biology_b,pds_biology_c}, to study the structure of proteins and
their complexes, and how these structures change during protein functioning.
Measuring a few distances in a protein cannot be compared with
obtaining the entire protein structure using methods such as X-ray
diffraction~\cite{xray_a,xray_b} or cryo-electron microscopy (cryo-EM)~\cite{cryoem_a,cryoem_b}, but EPR
spectroscopy has its advantages as a tool for refining and expanding data from
these two methods. First, it yields distance distributions, which enables
the study of mobile proteins that may adopt multiple conformations in the
sample, unlike X-ray and cryo-EM, for which mobility seriously interferes with
crystallization or with obtaining a clear image, respectively. Second, it is
environment-independent: it can work with any protein in any environment
(solvent, lipid membrane, micelles, etc.), since spin-coherence
characteristics do not depend on the interspin medium. Finally, PDS requires
relatively small amounts of protein; a typical sample volume for the
\SI{34}{GHz} spectrometer is \SI{5}{\micro\litre} and the minimum suitable
concentration is in the \SIrange{1}{10}{\micro M} range~\cite{deer_benchmark2021}.

The need for PDS methods is beginning to be recognized by the structural
biology community, since protein mobility is an integral part of their
functioning, and PDS can then add the necessary mobility information to the
knowledge of their 3D structure obtained from other methods or modeling~\cite{jumper2021}. On the other
hand, the interest and need for pulsed EPR spectrometers in the structural
biology community is held back by the cost of the spectrometers. Weighing
the cost against the extractable information, the cost side strongly outweighs
the information side for all biological laboratories at present. The aim of this work is to develop
cost-efficient equipment that provides an optimized scheme for the
entire spectrometer in terms of both sensitivity and cost.

Pulsed microwave (MW) bridges of EPR spectrometers form their pulses and detect
the EPR signal in one of three ways, distinguished by how the pulses are generated and,
decisively, by whether the working band is separated far enough from the local
oscillator (LO) to be filtered.

(i)~In pulse slicing, pulses are cut from a continuous source by fast MW
switches and phased with switched phase shifters, as in commercial Bruker
spectrometers and their modifications \cite{polyhach2012,schops2015}. Operating
with pulses of different phase requires either multiple channels with separate
phase shifters, as in the Bruker Elexsys, or a single fast phase shifter, as in
a custom-built spectrometer with a \SI{20}{ns} $0/90/180/270^\circ$
switch \cite{isaev2023}.

(ii)~In near-zero/unfiltered synthesis, a digital-to-analog converter
(DAC) synthesizes the pulses, but the working band is not separated from the LO
and its mirror and therefore cannot be filtered against them: either a
low-frequency DAC (output $<\SI{1}{GHz}$) places the signal near zero
frequency,~\cite{kaufmann2013,doll2019,isaev2023} or the synthesized band is
up-converted to the operating band with the LO and mirror bands left
unfiltered \cite{hill2023}. In the near-zero case the positive and negative
frequencies fold through zero, and detection at zero frequency without full
phase cycling increases experimental noise \cite{isaev2023}.

Both slicing and near-zero/unfiltered synthesis suffer from LO leakage wherever
an in-phase/quadrature (IQ) mixer is used. Regardless of the DAC resolution, an
analog IQ mixer always leaks residual LO power set by its finite LO--RF (radio
frequency) isolation, which no digital control can remove, so an attenuator is
required to minimize it.

(iii)~In intermediate-frequency (IF) synthesis, the pulses are
synthesized at an IF shifted from the source by more than \SI{1}{GHz}, with more
than \SI{1}{GHz} of bandwidth, far enough that the LO and mirror bands can be
filtered out \cite{doll2014,doll2017}. Such a bandwidth is typically enough to
work with different resonators and samples of different
permittivity \cite{jeschke_mlgr}. The
low-MW-power part then operates only as continuous-wave (CW) up- and
down-converters and needs no fast switches or phase shifters, which makes its
design price-efficient, and the leakage, amplitude-imbalance, and
multiple-pathway problems that constrain the first two approaches either vanish
or are easily corrected. Previously, this price gain could not compensate for
the more expensive set of necessary analog-to-digital converters (ADCs) and
DACs, but this article shows that this is no longer the case, and that systems
operating at up to \SI{5}{GHz}, \SI{10}{GS/s} effective sample rate and
more than \SI{1}{GHz} bandwidth can
reduce the final cost of a pulsed EPR spectrometer. Generation and detection
take place entirely at the intermediate frequency, so the same pulse-generation
and digitization electronics serve any operating band, such as
X\nobreakdash-, Q\nobreakdash-, or W\nobreakdash-band. Only the microwave bridge (the mixers, synthesizers, and filters for the
target range) is swapped.

Another problem is effective data transfer: since this IF spectrometer
requires control at much higher rates, it also involves large amounts of data input and
output compared to classic spectrometers, which calls for additional
optimization. This paper presents a hardware and software system, including a PCIe card
with a 16-bit DAC at a \SI{10}{GS/s} effective sample rate, a \SI{2.5}{GS/s}
14-bit ADC and an 11-channel
pulse generator on board, and its custom firmware which allows data input and
output in parallel with a running experiment. The system is integrated with the
Atomize open-source software~\cite{atomize,atomize_docs}, which altogether allows performing
pulsed EPR experiments with almost 100\% time efficiency, meaning no time is lost during the
writing of commands to the card or the reading of data from it. The system design, its firmware development, the details of
the high-level Python application programming interface (API), and its
performance in pulsed EPR applications are demonstrated. Although developed here
for pulsed EPR, the proposed hardware and software approach is not limited to it
and can be adapted to other experimental platforms and techniques that generate
very large volumes of data, where the efficient use of expensive measurement
time is a decisive factor.

\begin{figure*}[t]
  \centering
  \includegraphics[width=\textwidth]{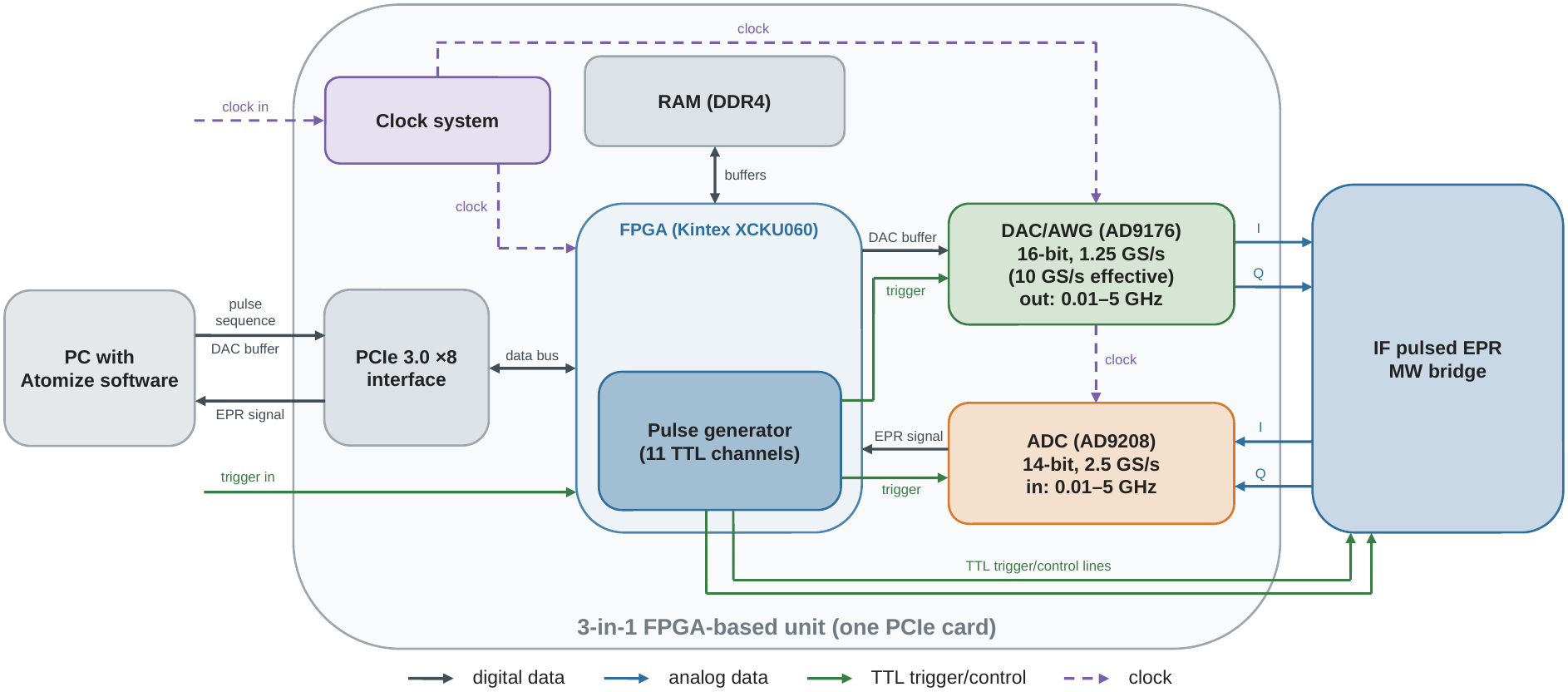}
  \caption{Block diagram of the integrated unit. A single PCIe card hosts three
    modules: the DAC/AWG, the ADC, and the multichannel TTL pulse generator
    implemented in the FPGA firmware. All three run from a common on-board
    reference clock. Arrows show the information flow: digital data exchanged
    with the host PC over the PCIe~3.0 $\times 8$ interface; analog I/Q signals exchanged
    with the IF pulsed EPR microwave bridge; and TTL lines from the pulse
    generator, which trigger the DAC, start the ADC, gate the high-power
    amplifier, and switch the low-noise amplifier protection in the bridge,
    keeping every path on a single on-board time base. The card also accepts an
    external reference at clock in and an external start signal at trigger in,
    so
    the unit can be locked to and triggered by other instruments.}
  \label{fig:block}
\end{figure*}

\section{System Design}
Figure~\ref{fig:block} shows the connections the card must provide. The
following input and output channels are required to operate the IF pulsed EPR
spectrometer: a DAC-generated
pulse input at IF, pulse high-power amplifier (HPA) control, low-noise amplifier (LNA) protection control, and an ADC input for the EPR
signal at IF. The design looks simple, because the low-power IF MW-bridge part operates in CW
mode, since no MW can penetrate the high-pass filter (HPF) without DAC pulses. A
single PCIe card with a fast DAC and ADC on board could therefore provide
all the control of the spectrometer. Since every card of this type has a
field-programmable gate array (FPGA)
and, in most cases, a digital signal connector, it allows implementation of a
multichannel transistor-transistor logic (TTL) pulse generator on its base through development of special
firmware. Because the DAC, ADC and pulse generator reside on the same board and
run from a single on-board reference clock, they are inherently synchronized:
there is no need to distribute and phase-lock separate clocks between the
generation and detection paths, as in multicard designs. There is no relative clock
drift and no jitter to degrade the phase coherence between excitation and detection or
the timing of the pulse generator.
An external reference input (clock in) allows the on-board
clock to be locked to another instrument, and an external trigger input
(trigger in) allows a pulse sequence to be started by an external event. All of
this is collected in Fig.~\ref{fig:block}: the three modules, the shared
on-board reference clock, and the flow of information between the card, the host
PC, and the IF pulsed EPR microwave bridge.

Integrating the three modules on a single card carries a higher one-time
development cost, but per card it is significantly cheaper in serial production.
Two such cards have already been built, and the savings relative to the six
separate cards used in our previous design have already offset that development
cost \cite{isaev2023}.

\begin{table}[!t]
\caption{Key Parameters of the Three On-Board Modules (DAC/AWG, ADC, and
  Multichannel Pulse Generator)}
\label{tab:specs}
\centering
\small
\begin{tabular}{@{}ll@{}}
\toprule
\multicolumn{2}{@{}l}{\textbf{DAC/AWG}}\\
\midrule
Chip                     & AD9176\\
Analog outputs           & 2, \SI{50}{\ohm}\textsuperscript{a}\\
Resolution               & 16 bit\\
I/Q data rate per output & \SI{1.25}{GS/s}\\
Core sample rate         & \SI{10}{GS/s} ($8\times$ interpolation)\\
Output amplitude         & $\pm\SI{0.26}{V}$ ($-1.7$\,dBm)\\
Output bandwidth         & \SIrange{0.01}{5}{GHz}\\
LO tuning grid           & \SI{312.5}{MHz}\\
Instantaneous bandwidth  & \SI{1.25}{GHz}\\
\midrule
\multicolumn{2}{@{}l}{\textbf{ADC}}\\
\midrule
Chip                     & AD9208\\
Analog inputs            & 2, \SI{50}{\ohm}\textsuperscript{b}\\
Resolution               & 14 bit\\
Sample rate per input    & \SI{2.5}{GS/s}\\
Input range              & $\pm\SI{1.5}{V}$\\
Input bandwidth          & \SIrange{0.01}{5}{GHz}\textsuperscript{c}\\
Sustained output rate    & \SI{6}{GB/s}\\
Registration window      & \SI{3.2}{ns}--\SI{50}{\micro s}\textsuperscript{d}\\
On-board averaging       & up to 65535\\
\midrule
\multicolumn{2}{@{}l}{\textbf{Pulse generator}}\\
\midrule
FPGA                            & Kintex UltraScale XCKU060\\
Core frequency                  & \SI{312.5}{MHz}\\
Time step                       & \SI{3.2}{ns}\\
Channels                        & 11\\
Pulse amplitude                 & $>\SI{2.8}{V}$ at \SI{330}{\ohm}\\
Pulse fronts                    & \SI{1.4}{ns}\\
Max.\ single time interval      & $\approx\SI{13.74}{s}$\\
Max.\ number of time intervals  & 622\\
Synchronization                 & external trigger and clock\\
\bottomrule
\end{tabular}

\vspace{3pt}
\begin{minipage}{\columnwidth}
\footnotesize
\textsuperscript{a}Two independent real RF outputs: the four \SI{1.25}{GS/s}
DAC cores of the AD9176 are paired into two complex (I/Q) channels, and on-chip
digital up-conversion turns each I/Q pair into one real RF output. Each output
is therefore programmed by a complex 16-bit waveform, and the two outputs can be
connected directly to the I and Q ports of an analog IQ mixer.\\
\textsuperscript{b}Two independent real inputs; in quadrature detection they
carry the I and Q outputs of the MW bridge, and the down-conversion of the
intermediate frequency is performed in software (Section~\ref{sec:software}).\\
\textsuperscript{c}Spans four Nyquist zones of the \SI{2.5}{GS/s} sampling; a
zone above the first is used in bandpass (undersampling) detection, with the MW
bridge restricting the input to the chosen zone (Section~\ref{sec:adc}).\\
\textsuperscript{d}Not a strict limit; a longer registration window increases
the data-processing overhead (see Section~\ref{sec:performance}).
\end{minipage}
\end{table}

\subsection{DAC/AWG}
The DAC operates as an arbitrary waveform generator (AWG), and its
performance was optimized for pulsed EPR needs, which do not
require more than a \SI{1.25}{GHz} frequency sweep in one pulse in most
experiments. The AD9176 chip (Analog Devices, USA) is a 4-channel DAC with a
real data rate of \SI{1.25}{GS/s} per channel. Its four channels are combined
into two complex outputs: $8\times$ on-chip interpolation brings the I and Q
streams up to the \SI{10}{GS/s} core sample rate of the DAC, i.e.,\ a Nyquist
range of \SI{5}{GHz}, and a
digital IQ mixer driven by an internal LO frequency generator then shifts the
signal to the desired frequency. This on-chip up-conversion functions without
the inherent LO leakage of analog MW mixers, so the output can be shifted
anywhere within the \SIrange{0.01}{5}{GHz} output band. The LO is set from 0 to
\SI{5}{GHz} on a grid of \SI{312.5}{MHz}; frequencies between grid points are
reached with the I/Q waveform itself, which at the \SI{1.25}{GS/s} data rate
spans $\pm\SI{625}{MHz}$ around the LO, so the coarse LO grid does not restrict
the accessible frequencies. As with the MW IQ mixer, one
has to define the LO frequency and write the data to the I and Q channels to
obtain the desired RF output.

The pulse-position step is set by the
\SI{1.25}{GS/s} data rate and equals \SI{0.8}{ns}, which is sufficient for all
pulsed EPR experiments; the DAC parameters are summarized in Table~\ref{tab:specs}. The DAC is
started by an internal trigger from the pulse generator, transmitted over the
card bus, and the trigger length determines
the length of the DAC output signal.

Every pulse in an output sequence is an independent entity: its waveform is synthesized
separately, stored in the DAC buffer, and played back by its own trigger from
the pulse generator, so that shape, length, intermediate frequency, phase, and
amplitude are all set pulse by pulse (Section~\ref{sec:software}). They may differ
between neighboring pulses of the same shot as much as between successive
shots, within the $\pm\SI{625}{MHz}$ that the \SI{1.25}{GS/s} I/Q data rate
spans around the digital LO, which is fixed for the whole sequence.

The output amplitude corresponds to
\SI{$-1.7$}{dBm}, which lies in
the linear regime of IQ mixers like the Marki MMIQ-0218L (Marki Microwave, USA), whose up-conversion
input $P_{1\mathrm{dB}}$ is \SI{2}{dBm}. The DAC output can therefore be connected to
the mixer I and Q inputs directly. Thus, the choice of such a DAC chip is
cost-effective and supports pulsed EPR experiments at intermediate
frequencies with close to optimal parameters.

\subsection{ADC}\label{sec:adc}
The ADC chip AD9208 (Analog Devices, USA) is specified in Table~\ref{tab:specs}. It is started by an internal
trigger from the pulse generator, transmitted over the card bus; the trigger
length sets the digitization window
(the registration duration is unlimited), and it supports on-board
averaging up to 65535 times. The performance of this ADC together with the input-bandwidth parameters allows
working in any of the four Nyquist zones: 0.01--1.25, 1.25--2.5,
2.5--3.75, and \SIrange{3.75}{5}{GHz}.
Operating in a higher zone is a form of bandpass (undersampling) detection, an
approach used earlier in EPR for time-locked subsampling~\cite{hyde1998}: the
analog bandpass filtering in the MW bridge restricts the input to the chosen
zone, so that noise from the other zones is not aliased into the digitized band. The
fourth zone enables effective filtering of the LO and mirror bands in both the
receiver and the transmitter, which allows the LO to be tuned over a \SI{3}{GHz}
range. Shifting the middle of the \SI{1.25}{GHz} observation/generation window
accordingly provides more than \SI{4}{GHz} of IF spectrometer bandwidth, enough
to satisfy the vast majority of pulsed experiment requirements.

\subsection{Pulse Generator}
The pulse generator was developed on the base of a Kintex UltraScale XCKU060
FPGA (AMD/Xilinx, USA) used in the InSys 126P card; its parameters are listed in
Table~\ref{tab:specs}. The 32-bit hold-time field is clocked at \SI{312.5}{MHz}
(\SI{1.25}{GHz}/4, coherent with the DAC; $\SI{3.2}{ns}$ per tick), so a single time interval can be held from
$\SI{3.2}{ns}$ to $2^{32}\times\SI{3.2}{ns}\approx\SI{13.74}{s}$. Because the pulse generator merely triggers the DAC and ADC and activates the
HPA and LNA protection in the IF pulsed MW bridge, its \SI{3.2}{ns} time
resolution is fully adequate for these tasks.

\section{Firmware}
The goal of this work was to make the three modules (ADC, DAC and pulse
generator) operate pulsed EPR experiments with 100\% time efficiency. This
means that there is no overhead for data input/output to run the next pulse
sequence with fully independent AWG pulse positions and shapes, which gives complete
freedom to conduct the pulsed EPR experiment.

This freedom was implemented in the spectrometer of \cite{isaev2023}, where
it enabled nonuniform sampling and fully symmetrical phase cycling (4 phases for
each pulse) for fixed delays in the AWG pulse sequence before moving on to the
next delays. The ability to start another AWG sequence greatly helps to tune AWG
experiments, since real-time phase cycling with AWG pulses can be performed while
tuning. With the AWG echo spectrum available,
the resonator dip position, the AWG pulse frequency and the magnetic field could
easily be adjusted. The problem of that earlier design was the use
of three different PCI/PCIe cards (ADC, DAC and pulse generator) which did not know
about each other, so starting a new AWG
pulse sequence required several steps with time penalties: calculating the pulse program and
loading it into the pulse generator (not less than \SI{1}{ms}); calculating the
AWG program and loading it into the DAC card (not less than \SI{1}{ms}); the ADC
measurement; transferring the data (not less than \SI{0.1}{ms}); and simple
mathematical operations (usually \SIrange{1}{10}{\micro s}, which can be done in
a parallel thread). Moving on to the next measurement point therefore took about
\SIrange{2}{5}{ms}. This overhead is
negligible when each pulse sequence is accumulated for several tens or even
hundreds of milliseconds before the next one is started. When the accumulation
time per sequence is reduced to on the order of \SI{10}{ms}, however, the overhead
becomes a substantial fraction of the total experiment time. Such short
per-sequence accumulation is often unavoidable for experiments with a large
phase cycle or a large number of measurement points: accumulating longer would
stretch a single scan to an hour or more, and since the hardware can drift
slightly over such timescales, the data must instead be collected by repeating
short scans many times. Eliminating this overhead in IF spectrometers is
therefore the principal goal of the present design.

\subsection{Firmware Operation With the Pulse Generator Module}
\label{sec:pulser_fw}
In order to provide continuous operation of the pulse generator, it is necessary
to record the next pulse sequence while the current one is operating. This is
achieved by having two independent FIFO (first in, first out) buffers, one of which is being executed
while the other is simultaneously rewritten/reflashed.

To ensure that information is delivered without delay, we chose the nonburst 
data transfer mode, which bypasses standard PCIe data packeting and
writes data directly to the desired memory location. The speed of
such recording is slower, at around \SI{100}{MB/s} (or \SI{100}{kB/ms}).
Although slower than the maximum speed of \SI{8}{GB/s} for the PCIe~3.0
$\times 8$ on which this card operates, it is quite sufficient for filling the
pulse-sequence buffers.

\begin{figure}[t]
  \includegraphics[width=\columnwidth]{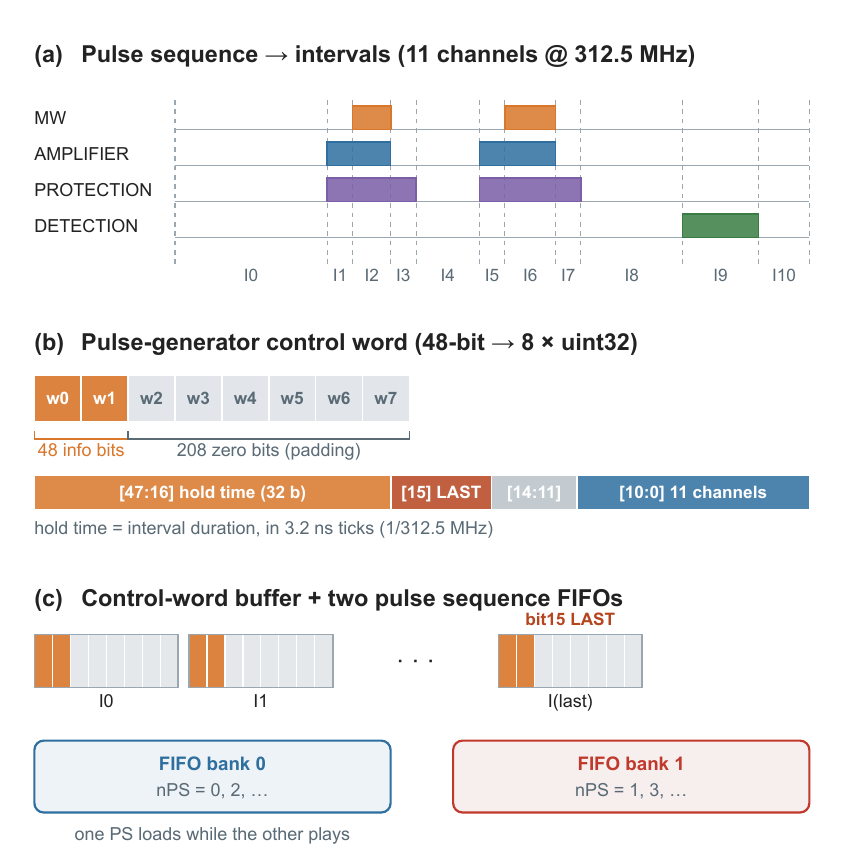}
  \caption{Multichannel TTL pulse-generator buffer. (a)~A pulse sequence is
    sliced into constant-output time intervals on the pulse-generator channels;
    the MW, amplifier, protection and detection channels shown here are only an
    example, and other control channels (e.g.,\ AWG, laser triggers) can be
    defined in the same way. (b)~Each interval is encoded as a 48-bit
    pulse-generator control word (32-bit hold time, the LAST flag that marks the
    final interval of the sequence, and the 11 channel bits) padded to
    $8\times$\,uint32. (c)~The control-word buffer feeds two pulse-sequence
    FIFOs: one is played while the other is reloaded.}
  \label{fig:pulser}
\end{figure}

The pulse sequences are represented as consecutive time
intervals that contain information about which channel should be activated and
for how long. A time interval has 48 bits: 11 for channels, 5 unused,
and 32 to set the duration of that specified state of the
outputs (Fig.~\ref{fig:pulser}). This time interval is further expanded to
256 bits and transmitted to the card. The maximum number of time intervals is
622. In the worst case, when none of the pulse edges coincide, each pulse consumes two
time intervals (its own duration and the following gap), which corresponds to a
maximum of 311 pulses per sequence, counted across all channels.

The execution of pulse sequences occurs in a cyclic mode; there is no
waiting mode between pulse sequences. At the start of the experiment, the first
pulse sequence is written into one buffer and execution begins immediately; the
second sequence is written into the other buffer while the first is already
running. The
execution of the first buffer and the seamless transition to the second buffer
raises a flag signaling readiness to rewrite the first buffer, which starts the
rewriting process. Another flag is raised upon completion of writing to the
nonexecuting buffer. A final flag is read after an execution sequence is
complete. If the final flag is raised, execution of the pulse sequence switches
seamlessly to the other buffer. Otherwise, a repeated measurement of the current
sequence is performed. Each pulse sequence contains its own identifier, so that
repeated and consecutive measurement steps are never mixed up
[see also Fig.~\ref{fig:adc}(b)].

\subsection{Firmware Operation With the DAC/AWG Module}
The DAC module is handled in exactly the same way: two independent FIFO buffers,
one of which is played while the other is reloaded, so that the waveform for the
next pulse sequence is written while the current sequence is being executed. In
addition to the pulse-generator words, the host writes the DAC waveform buffers
themselves to the card at the same time.

\begin{figure}[t]
  \includegraphics[width=\columnwidth]{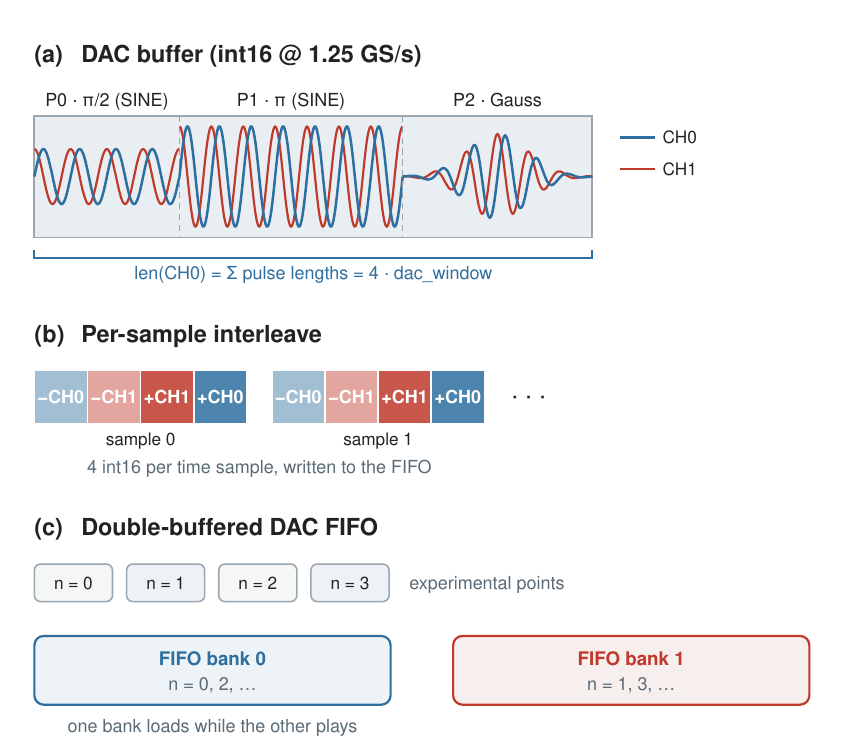}
  \caption{DAC/AWG buffer. (a)~The two int16 sample arrays of the AWG output
    channels CH0 and CH1 (\SI{1.25}{GS/s}), with the AWG pulses packed
    back-to-back; for pulsed EPR, CH1 is generated from CH0 with a
    $90^\circ$ phase shift. (b)~Four interleaved int16 values are written
    to the FIFO per time sample: with the two arrays in quadrature they supply
    the I and Q parts of both channels ($-$CH0, $-$CH1, $+$CH1, $+$CH0).
    (c)~The two FIFO banks are reloaded per experimental point: one bank loads
    while the other plays.}
  \label{fig:dac}
\end{figure}

The DAC array for each channel has 16-bit complex data: 16 bits
for the I and 16 bits for the Q part of the data. Each time sample occupies
32 bits for one channel or 64 bits for two channels. The relative phase of the
two output channels is set by the user; for the pulsed EPR experiments reported
here it is $90^\circ$, so that the same two arrays also provide the I and Q
parts of both channels. The array spans the
combined length of all DAC pulses in the sequence, which are packed back-to-back
(Fig.~\ref{fig:dac}). These two \SI{1.25}{GS/s} I/Q
channels are up-converted on-chip by the digital IQ mixer into two RF outputs at
the \SI{10}{GS/s} core sample rate. The length of the DAC trigger
pulse, generated by the pulse generator, determines the length of the DAC
output signal. The DAC buffer holds the entire pulse sequence, so each pulse-generator trigger
plays back only the corresponding pulse from it. The buffer is cyclic, so that
the repeated shots used for signal averaging reuse it without reprogramming.

\subsection{Firmware Operation With the ADC Module}
The ADC can be triggered by the pulse generator only once per pulse sequence,
and can digitize any number of data points, determined by the length of the trigger
pulse. The only limit to the amount of digitized data is that the raw data
flow must not exceed the rate at which the card can stream it to the host. The
ADC module has two 14-bit channels with a sampling rate of \SI{2.5}{GS/s}, and
each sample of each channel is stored as a 32-bit integer, so during acquisition
the raw data flow reaches \SI{20}{GB/s}. This exceeds both the \SI{8}{GB/s} theoretical
bandwidth of the PCIe~3.0 $\times 8$ interface and the \SI{6}{GB/s} rate that the
card sustains in practice (Table~\ref{tab:specs}). Thus, it is not possible to
run fully continuous experiments, but the card allows digitizing
${\approx}\,30\%$ of all time.

\begin{figure}[t]
  \includegraphics[width=\columnwidth]{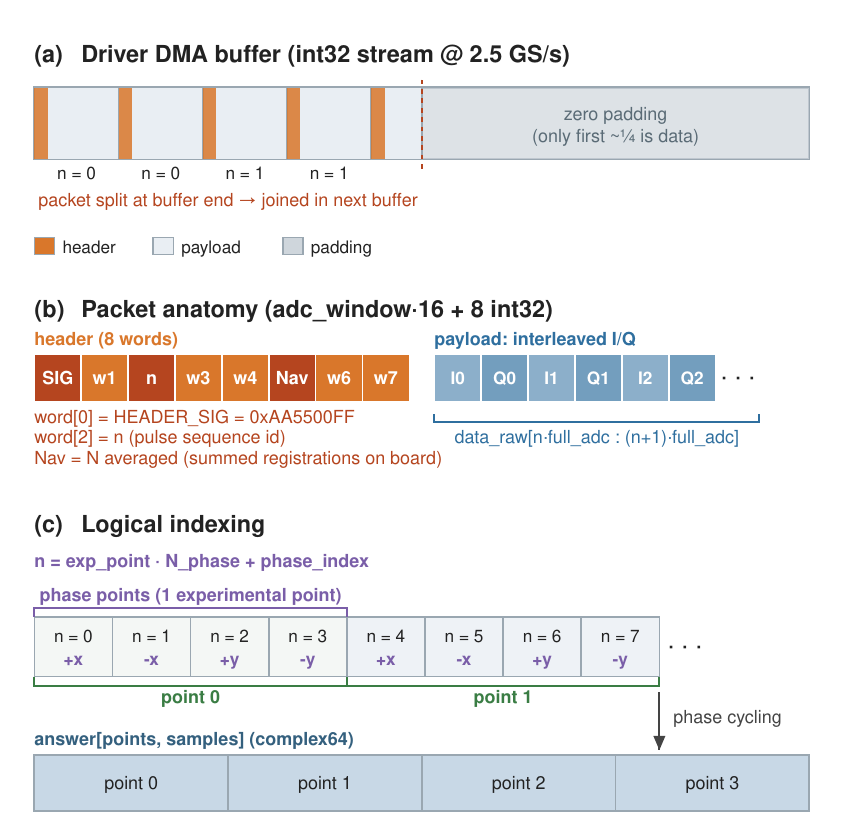}
  \caption{ADC streaming buffer. (a)~The driver direct memory access (DMA) buffer is a flat int32
    stream of back-to-back packets (only the first quarter holds data).
    (b)~Packet anatomy: an 8-word header (signature, pulse-sequence id $n$,
    number of averages) followed by the interleaved I/Q payload. (c)~Logical
    indexing assigns each packet the index $n =
    \text{exp\_point}\cdot N_{\text{phase}} + \text{phase\_index}$; phase cycling
    then combines the $N_{\text{phase}}$ packets of each experimental point into
    the final \texttt{answer[points, samples]} array.}
  \label{fig:adc}
\end{figure}

The ADC has a buffer of 64 units that are circularly rewritten. The size of each
unit can be controlled by the user (from \SI{32}{KiB} up to \SI{4096}{KiB}). A
quarter of this buffer is occupied by digitized data; the rest is filled with
zeros (Fig.~\ref{fig:adc}). Each digitized window also has a header with the
pulse-sequence number, the measurement number, the number of on-board
averages, etc. This ensures that their assignment to the pulse sequence is
unambiguously interpreted. Thus, even if the user selects a suboptimal run time
for one pulse sequence that is not sufficient to reflash the second buffer, the
spectrometer will operate at 100\% efficiency by making several repeated
measurements of that pulse sequence. The data will be provided to the
user, who can decide whether to use it to enhance signal averaging or to discard
it.

\begin{table*}[!tp]
\caption{Most Important Functions and Their Main Arguments in the High-Level
  Python API, Grouped by On-Board Module}
\label{tab:api}
\centering
\small
\begin{tabular}{@{}%
  >{\raggedright\arraybackslash}p{0.23\textwidth}%
  >{\raggedright\arraybackslash}p{0.31\textwidth}%
  >{\raggedright\arraybackslash}p{0.39\textwidth}@{}}
\toprule
Function & Main arguments & Purpose\\
\midrule
\multicolumn{3}{@{}l}{\textbf{Pulse generator}}\\[3pt]
\texttt{pulser\_pulse} & \texttt{'channel'}, \texttt{'start'}, \texttt{'length'}, \texttt{'delta\_start'}, \texttt{'length\_increment'}, \texttt{'phase\_list'} & Define a TTL pulse: channel, position, duration, per-step increments, phase program\\
\texttt{pulser\_shift} / \texttt{pulser\_increment} & pulse names & Step pulse positions / lengths during a sweep\\
\texttt{pulser\_next\_phase} & --- & Advance the pulse generator to the next entry of~\texttt{'phase\_list'} (phase cycling)\\
\texttt{pulser\_update} & --- & Write the pulse sequence to the generator and start it (without phase cycling)\\
\texttt{pulser\_pulse\_reset} & pulse names & Restore pulses to their defined positions and lengths\\
\texttt{pulser\_repetition\_\allowbreak rate} & rate & Set the shot repetition rate\\
\midrule
\multicolumn{3}{@{}l}{\textbf{DAC/AWG}}\\[3pt]
\texttt{awg\_pulse} & \texttt{'func'}, \texttt{'frequency'}, \texttt{'amplitude'}, \texttt{'start'}, \texttt{'length'}, \texttt{'delta\_start'}, \texttt{'length\_increment'}, \texttt{'phase\_list'} & Define an AWG pulse: shape, intermediate frequency, amplitude, position, duration, per-step increments, phase program, etc.\\
\texttt{awg\_shift} / \texttt{awg\_increment} & pulse names & Step AWG pulse positions / lengths during a sweep\\
\texttt{awg\_next\_phase} & --- & Advance the AWG to the next entry of \texttt{'phase\_list'} (phase cycling)\\
\texttt{awg\_update} & --- & Load and start the AWG output (without phase cycling)\\
\texttt{awg\_pulse\_reset} & pulse names & Restore AWG pulses to their defined positions and lengths\\
\texttt{awg\_correction} & \texttt{'model'}, \texttt{'phase\_correction'}, \texttt{'f0'}, \texttt{'q\_factor'} & Resonator-profile predistortion (amplitude and phase) of swept AWG pulses\\
\midrule
\multicolumn{3}{@{}l}{\textbf{ADC}}\\[3pt]
\texttt{digitizer\_get\_curve} & \texttt{'partial'}, \texttt{'integral'}, \texttt{'skip\_redundant'} & Acquire and phase-cycle the signal; \texttt{'partial'} returns only the points updated since the previous call (for incremental plotting), with optional integration in the API and dropping of redundant repetitions\\
\texttt{digitizer\_at\_exit} & \texttt{'integral'} & Final readout at the end of the experiment (and in Stop handlers): waits for the background processing to finish and returns the exact accumulated arrays, full or integrated\\
\texttt{digitizer\_demodulate} & I and Q arrays, \texttt{'freq'}, \texttt{'ph'}, \texttt{'ph1'}, \texttt{'ph2'}, \texttt{'integral'} & Digital down-conversion of the acquired traces: demodulate the I and Q quadratures at the (negative) AWG pulse frequency \texttt{'freq'}, apply zero-, first- and second-order phase corrections, and return the full demodulated trace or its integral over the detection window\\
\texttt{digitizer\_number\_\allowbreak of\_averages} & number & Set the number of averages per phase\\
\texttt{digitizer\_decimation} & factor & Reduce the effective sample rate for long detection windows (applied on the fly while the stream is accumulated)\\
\texttt{digitizer\_processing\_\allowbreak thread} & \texttt{0/1} & Background processing of the ADC stream (enabled by default): buffers are averaged, normalized and phase-combined in a worker thread that overlaps the acquisition\\
\texttt{digitizer\_expand\_\allowbreak phase\_cycling} & per-pulse phase programs & Expand the compact (bracket) phase notation into the full phase table and derive the detection phase from the coherence-transfer pathway\\
\bottomrule
\end{tabular}
\end{table*}

\subsection{Firmware Operation With the High-Level Python Software}\label{sec:software}
The low-level firmware has an API written in the C programming language. The
API has all the functions necessary to efficiently operate the unit and
transfer the data from the internal memory to the PC. In order to further
simplify interaction with the card, we created a high-level Python module for
the Atomize software~\cite{atomize} using \texttt{ctypes}. The Python module
consists of three parts: the pulse generator, DAC, and ADC. In the
pulse-generator part of the module, pulse sequences are converted internally to
the firmware format from user-friendly input. The latter is self-explanatory to
pulsed EPR experimentalists and is based on several keyword arguments of the
function \texttt{pulser\_pulse}, namely \texttt{'channel'}, \texttt{'start'},
\texttt{'length'}, \texttt{'delta\_start'}, \texttt{'length\_increment'} and
\texttt{'phase\_list'}. Phase programs are specified through \texttt{'phase\_list'}
either explicitly (\texttt{'+x'}, \texttt{'-x'}, \texttt{'+y'}, \texttt{'-y'}) or
in the compact bracket notation~\cite{stoll2008}
(\texttt{[x]}, \texttt{(x)}, etc.); the \texttt{digitizer\_expand\_phase\_cycling}
function expands the compact form into the full phase table and computes the
corresponding detection phase from the selected coherence-transfer pathway. The user can change the positions, phases, pulse
durations and repetition rate of the pulse sequence using several specialized
functions: \texttt{pulser\_shift}, \texttt{pulser\_increment},
\texttt{pulser\_repetition\_rate}, etc. The progress of the experiment is
controlled by the functions \texttt{pulser\_next\_phase},
\texttt{pulser\_update} and \texttt{pulser\_pulse\_reset}. The most important
functions and their main arguments are summarized in Table~\ref{tab:api}.

The DAC part of the high-level module generally mimics the pulse generator.
There is an \texttt{awg\_pulse} function with several additional keywords
required to describe DAC pulses: \texttt{'func'}, \texttt{'frequency'},
\texttt{'amplitude'}, etc. The \texttt{'func'} keyword sets the pulse shape:
constant-amplitude, Gaussian, sinc, WURST and sech/tanh
envelopes are available, the last two being frequency-swept (adiabatic) pulses.
For these swept pulses, the \texttt{awg\_correction} function applies a
resonator-profile predistortion that compensates the amplitude and phase
response of the resonator across the swept band. Positions, durations and increments of the DAC
pulses are controlled by corresponding trigger pulses generated by the
multichannel pulse generator. The phase of the pulses is controlled by the
keyword \texttt{'phase\_list'}, and phase cycling can be initiated by the
\texttt{awg\_next\_phase} function. Without phase cycling, the DAC module can be
launched by the \texttt{awg\_update} function. Overlapping AWG pulses are
supported: when pulses share a time interval their waveforms are summed
sample-by-sample into the same array and clipped to the 16-bit DAC range, as
required for composite- and shaped-pulse schemes. Two independent FIFO buffers
are handled automatically for both the pulse generator and the DAC module, and
the DAC buffer is not recomputed from scratch on every step: each pulse
waveform is cached against its defining parameters (shape, length, frequency,
phase, amplitude, etc.), so when the next experimental point only changes the
pulse timing (that is, it moves the pulse-generator triggers in time while
leaving every pulse shape unchanged), the cached samples are reused and the
time-expensive waveform synthesis is skipped. DAC buffers are processed using NumPy
arrays~\cite{numpy} and converted to a format suitable for the low-level
firmware using \texttt{ctypes}.

\begin{figure}[t]
\begin{lstlisting}[style=pyfig]
for k in general.scans(SCANS):
    for j in range(POINTS):
        for i in range(PHASES):

            pb.awg_next_phase()
            pb.pulser_update()

            a, b, rng = pb.digitizer_get_curve(
                POINTS, PHASES,
                current_scan = k, total_scan = SCANS,
                partial = True)

            if a is not None:
                i0, i1 = rng
                data[0][:, i0:i1] = a
                data[1][:, i0:i1] = b
                general.update_2d(
                    EXP_NAME, data, i0, i1, ...)

        pb.pulser_shift()
        pb.awg_shift()

    pb.pulser_pulse_reset()
    pb.awg_pulse_reset()
\end{lstlisting}
  \caption{Python code required to perform a general pulsed EPR experiment with
    AWG pulses and phase cycling. The definition of the pulses and other
    parameters, such as the array \texttt{data} for storing the results and the
    numbers of points and phases, are omitted for clarity. Called with
    \texttt{partial=True}, \texttt{digitizer\_get\_curve} returns the two signal
    quadratures for only the measurement points completed since the previous
    call, together with their index range \texttt{rng}, unpacked as \texttt{i0,
    i1} (the half-open point range \texttt{[i0:i1)} of \texttt{data} that
    changed). It returns \texttt{None} when no new ADC buffer has arrived, and
    that iteration is skipped. Those points are written into the persistent 2D
    array \texttt{data} and sent to the live plot with \texttt{update\_2d}, which
    redraws only \texttt{[i0:i1)} rather than the whole frame, so the per-readout
    cost scales with the size of the update instead of the size of the array.}
  \label{fig:code}
\end{figure}

The ADC part of the module has two main functions, namely
\texttt{digitizer\_get\_curve} and \texttt{digitizer\_number\_of\_averages}.
The second function is simple and straightforward by its name, while the
purpose and implementation details of the first are quite complex. To ensure
maximum efficiency of PCIe-bus usage, data from the ADC are transferred only
when at least one of the 64 buffer units is filled. A single unit usually holds
the data of several experimental points, so the
\texttt{digitizer\_get\_curve} function either transfers the data of a filled
unit from the buffer to the PC or simply returns without action if no unit is
filled yet. The data are transferred to the Python
module using the \texttt{numpy.frombuffer} function. After that, using the
pulse-sequence number from the ADC data header, the number of points, and the
number of phases in the experiment specified as function arguments, the
received array is processed into a final array corresponding to the quantity
measured in the pulsed experiment. If there is phase cycling in the experiment,
the function also performs it, according to the detection-phase array given by
the keyword \texttt{'acq\_cycle'}. By design, all of this processing (parsing
the packet stream, accumulation of the running average, normalization, and
phase cycling) is carried out in a background thread:
\texttt{digitizer\_get\_curve} only hands the transferred units to that thread
and immediately returns the measurement points completed since the previous
call, so the data handling overlaps the acquisition instead of delaying it
(Section~\ref{sec:performance}). The background processing can be switched off
with the \texttt{digitizer\_processing\_thread} function. At the end of the
experiment the final data are read
out with \texttt{digitizer\_at\_exit}, which waits for the processing thread to
finish the queued units and recomputes the accumulated arrays, so the saved
data are always complete on every exit path. Apart from the final data, it is also
possible to explicitly see the number of repetitions of each point in the pulse
sequence. Due to the design of the card, these numbers may exceed the number of
averages specified in the \texttt{digitizer\_number\_of\_averages} function by
an integer factor. Such extra repetitions occur when the next pulse sequence
cannot be written to the card in time and the current one is repeated instead
(Section~\ref{sec:pulser_fw}), i.e.,\ when the data-acquisition time per point of the pulse
sequence is less than the $\sim$\SI{2}{ms} on-the-fly reprogramming time. There is also a special regime of the
\texttt{digitizer\_get\_curve} function in which an integration of the data in
a given window is performed. Finally, when AWG pulses are used, the EPR signal is sampled
at the intermediate frequency, so the companion function \texttt{digitizer\_demodulate}
performs the digital down-conversion: it demodulates the I and Q traces at the
(negative) AWG pulse frequency, applies zero-, first- and second-order phase
corrections, and, like the integration regime above, can return either
the full demodulated trace or its integral over the detection window.
More general information about how the
module is designed in the Atomize software can be found
elsewhere~\cite{atomize,atomize_docs}.
The high-level module described above allows for a~clear and concise
description of the pulsed experiment, given in Fig.~\ref{fig:code}. The figure
shows the Python code required to perform a general pulsed EPR experiment with
AWG pulses and phase cycling. For clarity, the definition of the pulses and
other parameters is omitted. A detailed example of a complete experimental
script, a standard four-pulse DEER/PELDOR (double electron--electron resonance
/ pulsed electron--electron double resonance) measurement, is provided as supplementary
material; the benchmarks of Section~\ref{sec:performance} were run with the same
script by varying the parameters listed there.

\begin{figure*}[t]
  \centering
  \includegraphics[width=0.9\textwidth]{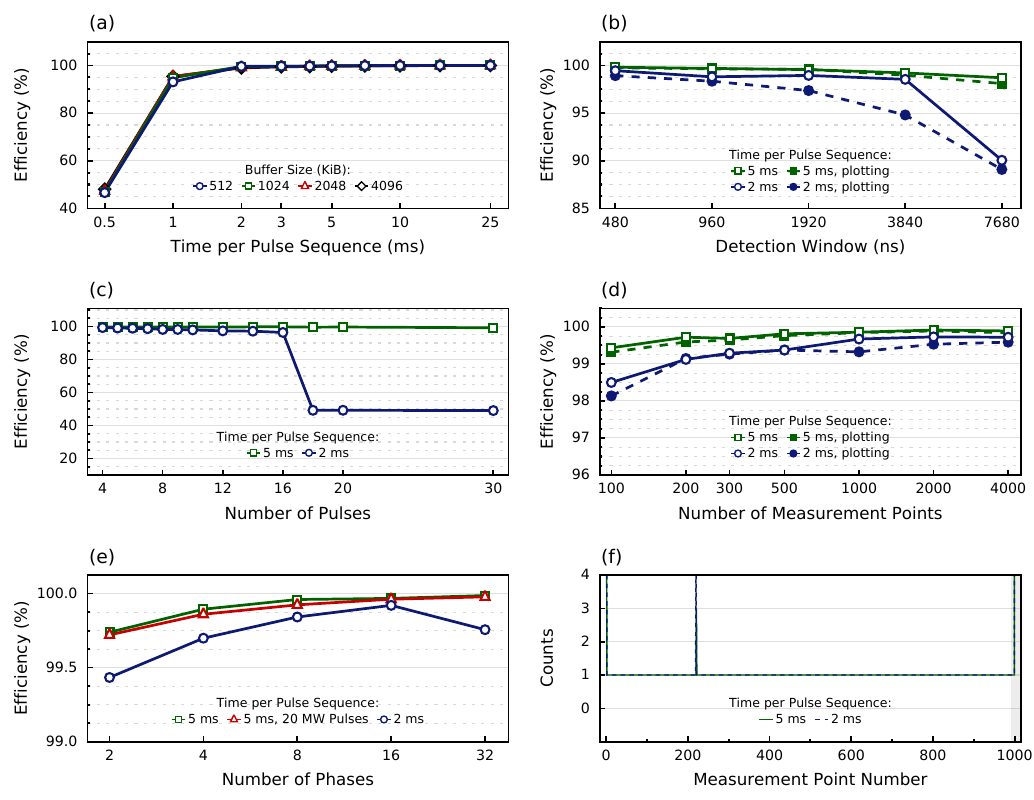}
  \caption{Time efficiency of the integrated unit, benchmarked with a four-pulse
    DEER sequence. (a)--(e)~Efficiency
    ($t_{\text{est}}/t_{\text{meas}}\times100\%$) versus (a)~the acquisition time
    per pulse sequence for several ADC buffer sizes, (b)~the detection-window
    length, (c)~the number of pulses, (d)~the number of measurement points, and
    (e)~the number of phases. The benchmarks in (a)--(c), (e), and~(f) were performed
    with live data plotting enabled, as in routine operation; panels~(b) and~(d)
    additionally isolate its cost, comparing the curves labeled ``plotting'' in
    the legend, in which the display is updated on each new ADC buffer arrival
    (filled symbols, dashed lines), with the otherwise identical scans where
    plotting is suppressed (open symbols, solid lines). The two nearly coincide,
    so live plotting adds only marginal overhead. 
    (f)~Number of counts recorded at each measurement point during a single scan
    at \SI{2}{ms} and \SI{5}{ms} per pulse sequence. For clarity, the individual steps
    of the phase cycle are here treated as separate measurement points. The
    final point (shaded) carries the redundant buffer-fill repetitions and is
    excluded from the efficiency in (a)--(e).}
  \label{fig:efficiency}
\end{figure*}

\section{Overall System Performance}\label{sec:performance}
The time efficiency of the integrated unit was characterized directly in a
typical pulsed EPR experiment. All benchmarks were run on a workstation with a
32-core AMD Ryzen Threadripper 3970X processor and \SI{128}{GiB} of RAM, running
Ubuntu~22.04.3 LTS (Linux kernel 6.5.0). The benchmark was built around a four-pulse
DEER experiment, run on the spectrometer from
\cite{borodulina2025}. All pulses were synthesized by the on-board DAC at
intermediate frequency and sine-shaped: the three observer pulses at
\SI{130}{MHz}, each \SI{44.8}{ns} long, with the $\pi/2$ pulse set to half the
amplitude of the two $\pi$ pulses, and a \SI{16}{ns} pump $\pi$ pulse at
\SI{200}{MHz}. The DEER trace was acquired by stepping the position of the pump
pulse in \SI{3.2}{ns} increments over 500 measurement points, using a
\SI{480}{ns} detection window (\SI{0.4}{ns} step), a simple two-step phase cycle, and a
\SI{1}{kHz} repetition rate. Throughout, the entire detection window, i.e.,\ the
so-called direct dimension~\cite{bowman2021}, is transferred, processed, and
plotted, rather than reduced on the fly to a single integrated value. The
acquired data set is therefore intrinsically two-dimensional, spanning the
direct (detection-window) dimension against the indirect dimension swept by the
experiment, which makes the input/output load, and hence the efficiency
benchmarked below, correspondingly heavier than for integrated detection.
Starting from this baseline, each parameter that
governs the data input/output load --- the ADC buffer size,
the number of pulses in the sequence, the
detection-window length, the number of measurement points, and the number of
phases --- was varied in turn while the others were held fixed. These scans are
reported in the panels of Fig.~\ref{fig:efficiency}. The time per pulse sequence
[Fig.~\ref{fig:efficiency}(a)], shown for several ADC buffer sizes, was varied
either through the repetition rate (down to \SI{0.5}{ms}) or through the number
of averages. For example, averaging 
2 times at each step of the phase cycle at a \SI{1}{kHz} repetition rate corresponds
to \SI{2}{ms} per pulse sequence. Each new phase triggers a full reprogramming of the card,
so each phase-cycle step is handled exactly like an entirely new
pulse sequence with arbitrary parameters and DAC waveforms. The dependence on sequence length [Fig.~\ref{fig:efficiency}(c)]
was probed by extending the observer block with additional refocusing pulses, 
up to 30 MW pulses. The total number of pulses created in the API is higher
than the number of MW pulses defined since each MW pulse has its own auto-generated 
protection and amplifier pulses. The efficiency is defined as the fraction of
the total experiment time ($t_{\text{meas}}$) that is
actually spent running pulse sequences: $\text{efficiency} =
t_{\text{est}}/t_{\text{meas}}\times100\%$, where $t_{\text{est}}$ is the
estimated time the card would need if every pulse sequence contributed to the
data with no dead time. In each benchmark, five identical scans were acquired and the
measured time was averaged over the scans, excluding the final measurement
point, whose buffer-fill overhead [Fig.~\ref{fig:efficiency}(f)] is a fixed
one-time cost rather than a steady-state contribution.

Figure~\ref{fig:efficiency}(a) shows the card efficiency as a function of the
time per pulse sequence, measured for several ADC buffer sizes. In this benchmark all pulse-sequence
parameters were kept constant, which makes it possible to estimate the time
needed to reprogram the card. At approximately \SI{1}{ms} per pulse sequence
the API begins to approach 100\% efficiency; the sharp drop below
\SI{1}{ms} occurs when a new pulse sequence cannot be written to memory in time,
so the current one is repeated instead (Section~\ref{sec:pulser_fw}). The measured efficiency
is essentially independent of the buffer size: every size tested reaches
99.5--99.9\% at \SI{5}{ms} per pulse sequence, and only near the \SI{1}{ms}
roll-off does a larger buffer help marginally, because it is processed less
frequently and is therefore less exposed to an occasional late sequence write.
The buffer size instead sets a trade-off that matters elsewhere: a larger buffer
is copied out of the driver less often, which lowers the per-buffer overhead that
becomes visible for long detection windows [Fig.~\ref{fig:efficiency}(b)], whereas
a smaller buffer is processed more frequently and refreshes the live display
faster, giving a more responsive data flow. A buffer of \SIrange{1024}{2048}{KiB}
balances the two for typical experiments, and at low ($<\SI{50}{Hz}$) and high
($>\SI{5000}{Hz}$) repetition rates the buffer size is adjusted automatically in
the API.

By design, the processing overhead of the high-level Python API is largely independent of the number of pulses in
the sequence or the number of detection phases. The benchmark confirms
this: at \SI{5}{ms} per pulse sequence the efficiency stays flat at 99.5\% as the number of pulses
[Fig.~\ref{fig:efficiency}(c)] and the number of phases
[Fig.~\ref{fig:efficiency}(e)] are varied, because the pulse-generator and DAC
buffers are reloaded in parallel with execution, and each ADC buffer has a fixed
size that is processed as soon as it is ready, independent of the number of
pulses or phases. The two-fold drop observed at \SI{2}{ms} per pulse sequence for 18 or more MW pulses is
again caused by repetition of the current pulse sequence (Section~\ref{sec:pulser_fw}), which
indicates that the reprogramming time of the card is close to \SI{2}{ms} for the parameters used here. Nevertheless,
\SI{5}{ms} per pulse sequence is sufficient to reach $>$99\% time efficiency even for 20 MW pulses
with a 32-step phase cycle, which is more than enough for pulsed EPR experiments
[Fig.~\ref{fig:efficiency}(e)].

On each ADC-buffer readout, the Python API updates and returns only the
measurement points completed since the previous readout, and the live display is
refreshed incrementally with just those points, so the per-readout cost is set by
the few points that changed rather than by the whole data set. The display is
handled off the measurement loop as well: the plotting call passes the new points
to a separate coalescing worker thread and returns within microseconds, and that
thread pushes only the latest update to the display, dropping any that is
superseded before it is drawn, so a slow display never restricts the acquisition. The
efficiency is therefore essentially independent of the number of measurement
points [Fig.~\ref{fig:efficiency}(d)], staying near 99\% from 100 to 4000
points, and the scans with live plotting (filled symbols, dashed lines) and with
plotting suppressed (open symbols, solid lines) nearly coincide. The small
remaining offset from 100\% behaves as a fixed cost rather than a per-point
one: the absolute overhead per scan stays near \SI{10}{ms} whether the scan
lasts \SI{0.4}{s} or \SI{16}{s}, and most of it is incurred once at the start
of the first scan, when the ADC stream is started, the accumulator arrays are
allocated, and the display is created. The slow rise of the efficiency with
the number of points [Fig.~\ref{fig:efficiency}(d)] and phases
[Fig.~\ref{fig:efficiency}(e)] is simply this constant overhead spread over an
increasingly long scan.

A weaker dependence remains on the detection-window length
[Fig.~\ref{fig:efficiency}(b)]. Unlike the number of pulses, phases, or
measurement points, the window length changes the volume of the digitized
data itself: every measurement point contributes one complex sample per
\SI{0.4}{ns} of the window, so at \SI{7.7}{\micro s} the card streams
sixteen times more raw data than at \SI{480}{ns}. The API is designed to
keep the handling of this stream off the acquisition timeline. The
per-sample work (summation into the running average, normalization, and
phase combination) is fused into a single pass over the data, and that pass
runs in a dedicated background thread which consumes each ADC buffer while
the card is already filling the next one. The readout call only hands the
arrived buffer to this thread and returns the measurement points completed
since the previous call. The one step that cannot leave the acquisition
loop is copying each \SI{2048}{KiB} buffer out of the driver, which takes a
few milliseconds. Following one buffer shows what this remainder means: at
a \SI{480}{ns} window one buffer holds about 210 pulse sequences and the card
needs \SI{420}{ms} (at \SI{2}{ms} per pulse sequence) to fill it, whereas at
\SI{7.7}{\micro s} the same buffer holds only 13 pulse sequences and is
refilled every $\approx$\SI{26}{ms}, so the same few-millisecond copy
consumes a sixteen times larger share of the timeline. The benchmark follows
this picture: at \SI{2}{ms} per pulse sequence the efficiency stays above
94\% up to \SI{3.8}{\micro s} and reaches 89--90\% at
\SI{7.7}{\micro s}, whereas at \SI{5}{ms} per pulse sequence it stays at or above
98\% over the whole range. For the detection windows
typical of pulsed EPR (below \SI{1}{\micro s}) the card thus runs at roughly 99\% time
efficiency. When an even longer window is required, the data volume can be
lowered at the source: the native \SI{0.4}{ns} sample spacing is usually
finer than required and can be coarsened with the
\texttt{digitizer\_decimation} function, which averages each group of
consecutive samples already while the stream is accumulated (acting as an
antialiasing filter, so out-of-band noise is not folded into the retained
band and the signal-to-noise ratio is preserved), or the window can be
integrated on the fly (the \texttt{'integral'} keyword of
\texttt{digitizer\_get\_curve}) to return a single value per point. The overhead
can also be reduced from the other side, by increasing the ADC buffer size up to
\SI{4096}{KiB}, so that each buffer accumulates more measurement points and the
fixed cost of copying a buffer out of the driver is paid a smaller share of the
acquisition time.

The scans with and without live plotting in Fig.~\ref{fig:efficiency}(b)
separate slightly only at \SI{2}{ms} per pulse sequence. The display itself never
paces the acquisition (the plotting call returns within microseconds, and
a coalescing worker thread pushes only the latest update); but that
thread shares the processor with the acquisition loop, and at \SI{2}{ms}
per pulse sequence, the reprogramming leaves almost no headroom, so
occasionally a delayed sequence write makes the card repeat the current
sequence (Section~\ref{sec:pulser_fw}), costing another whole repetition. At
\SI{5}{ms} per pulse sequence the same work fits into the available slack, and
plotting adds nothing at any window.

Finally, Fig.~\ref{fig:efficiency}(f) shows the number of counts recorded at
each measurement point over a single scan (whereas the efficiency in the other
panels is averaged over five), both at \SI{2}{ms} and \SI{5}{ms}
per pulse sequence, so that the ideal count is one per point. Since the buffer processing
is decoupled from the acquisition, the arrival of an ADC buffer does not
interrupt the pulse train, and the counts stay at the ideal value across
essentially the whole run, apart from isolated points near the start of the
scan. The only exception is the very
last point of the experiment, which always carries substantial redundant
acquisition, because the final ADC buffer on the card must be completely filled
before it can be transferred to the API. The Python API supports both options: (i) keeping the redundant
repetitions or (ii) discarding them, controlled by the
\texttt{'skip\_redundant'} keyword of the \texttt{digitizer\_get\_curve}
function. The former allows 100\% of the experimental time to improve the
measurement even for nonoptimal card parameters, since the redundant
repetitions still contribute to the signal averaging.

\section{Conclusion}
A single-board PCIe FPGA solution integrating a 16-bit DAC at a \SI{10}{GS/s}
effective sample rate, a
\SI{2.5}{GS/s} 14-bit ADC (both with up to \SI{5}{GHz} bandwidth),
and an 11-channel pulse generator has been presented
for pulsed EPR spectroscopy. The three modules share a common firmware that
records the next pulse sequence and reloads the DAC and pulse-generator buffers
while the current sequence is still running, and that streams the digitized data
over the PCIe bus through a custom transfer format. As a result, data input and
output proceed in parallel with the experiment, and virtually no measurement
time is lost to reading or writing.

The card is fully integrated into the open-source Atomize software through a
high-level Python API, so that a complete experiment is defined and controlled at
the level of pulse sequences rather than firmware. Benchmarking with a
four-pulse DEER sequence showed that the unit reaches close to 100\% time
efficiency for realistic conditions: an accumulation of about
\SI{5}{ms} per pulse sequence, limited by the $\sim$\SI{2}{ms} reprogramming time of the card, is
sufficient, and the efficiency is essentially independent of the other experimental
parameters, such as the number of pulses, the number of measurement points, and the
phase-cycling depth. The only residual cost scales with the number of ADC
samples acquired per point, i.e.,\ the detection-window length, and can be kept
negligible by decimation or on-the-fly integration in the Python API. Live plotting, offloaded to a separate coalescing worker thread,
adds essentially nothing, and even the redundant buffer-fill repetitions can be
retained to contribute to the signal averaging so that no measurement time is
wasted.

Because every pulse sequence is reprogrammed on the fly without dead time, the
number of averages and the positions of the measurement points can be varied
freely during the experiment, so that nonuniform sampling~\cite{isaev2023,matveeva2021}
and nonuniform acquisition~\cite{isaev2023,nekrasov2025} are supported natively
at no cost in time efficiency. More generally, since each acquisition step reprograms
the card in full, successive sequences may differ arbitrarily in their parameters
and DAC waveforms at no additional cost, leaving the experimenter free to tailor
the measurement protocol.

A further consequence of the seamless cyclic operation of the pulse generator
(Section~\ref{sec:pulser_fw}) is that the shot-to-shot timing stays constant throughout an
entire experiment: the card never halts the pulse train to reload and advances
to the next sequence only at a sequence boundary, so the repetition period is
preserved when an interpulse delay $\tau$, the phase-cycle step, or the magnetic
field $B_0$ is changed. This matters whenever the sample contains slowly
relaxing spins. When several paramagnetic species with overlapping spectra are
present, relaxing them fully would require pulsing, for example, at $<\SI{20}{Hz}$ instead of
the $>\SI{1}{kHz}$ allowed by the spins of interest, at a large cost in
sensitivity. Pulsing rapidly instead drives the slow spins into a steady state
in which they contribute only a constant background. On instruments that stop
and reload the pulse programmer at every step, this steady state is disturbed
after each change and the first few shots carry a variable, ill-defined
background; here it is perturbed only once, at the very first point of the
experiment, which can simply be discarded. The same deterministic behavior
benefits any measurement that must pulse rapidly to gain sensitivity,
experiments in which the pulse generator also triggers a laser at a fixed rate,
and field-swept experiments, where a constant interstep
timing keeps the average $B_0$ at each step identical on every scan.

Together, the high time efficiency and the on-the-fly flexibility described above
translate into a practical advantage. The proposed unit helps to lower one of the
barriers that have kept many structural-biology laboratories from routine access
to pulsed dipolar EPR spectroscopy, by making an intermediate-frequency architecture
both data-efficient and affordable.
Combining this 3-in-1 board with a price-optimized IF microwave bridge should
enable compact, cost-effective, and high-throughput pulsed EPR spectrometers.

\section*{Acknowledgment}
The development of the low-level firmware for the 3-in-1 FPGA unit and the development of the high-level Python API for the ADC and DAC were funded by the Russian Science Foundation, grant number 23-73-00042. A.R.M., A.S.I., and S.L.V. thank the Ministry of Science and Higher Education of the Russian Federation for granting access to the equipment for testing the high-level Python API. A.V.V. thanks the Ministry of Science and Higher Education of the Russian Federation within the governmental order for SRF SKIF Boreskov Institute of Catalysis for granting access to the equipment for developing the high-level Python API for the TTL pulse generator. The authors thank M.~K.~Bowman for ideas, useful discussions, and advice, as well as M.~V.~Fedin for his contribution at an early stage of the project.

\bibliographystyle{IEEEtran}
\bibliography{references,ieee_bstctl}

\begin{thebibliography}{10}
\providecommand{\url}[1]{#1}
\csname url@samestyle\endcsname
\providecommand{\newblock}{\relax}
\providecommand{\bibinfo}[2]{#2}
\providecommand{\BIBentrySTDinterwordspacing}{\spaceskip=0pt\relax}
\providecommand{\BIBentryALTinterwordstretchfactor}{4}
\providecommand{\BIBentryALTinterwordspacing}{\spaceskip=\fontdimen2\font plus
\BIBentryALTinterwordstretchfactor\fontdimen3\font minus
  \fontdimen4\font\relax}
\providecommand{\BIBforeignlanguage}[2]{{%
\expandafter\ifx\csname l@#1\endcsname\relax
\typeout{** WARNING: IEEEtran.bst: No hyphenation pattern has been}%
\typeout{** loaded for the language `#1'. Using the pattern for}%
\typeout{** the default language instead.}%
\else
\language=\csname l@#1\endcsname
\fi
#2}}
\providecommand{\BIBdecl}{\relax}
\BIBdecl

\bibitem{pds_review_a}
P.~E. Spindler, P.~Sch\"ops, A.~M. Bowen, B.~Endeward, and T.~F. Prisner,
  ``Shaped pulses in {EPR},'' \emph{eMagRes}, vol.~5, p. 1477, 2016.

\bibitem{pds_review_b}
G.~Jeschke, ``Dipolar spectroscopy -- double-resonance methods,''
  \emph{eMagRes}, vol.~5, p. 1459, 2016.

\bibitem{pds_review_c}
S.~Van~Doorslaer, ``Hyperfine spectroscopy: {ESEEM},'' \emph{eMagRes}, vol.~6,
  p.~51, 2017.

\bibitem{pds_biology_a}
R.~Dastvan and S.~Stoll, ``Recent advances in quantifying protein
  conformational ensembles with dipolar {EPR} spectroscopy,'' \emph{Curr. Opin.
  Struct. Biol.}, vol.~94, p. 103139, 2025.

\bibitem{pds_biology_b}
G.~Jeschke, ``The contribution of modern {EPR} to structural biology,''
  \emph{Emerg. Top. Life Sci.}, vol.~2, p.~9, 2018.

\bibitem{pds_biology_c}
D.~Goldfarb, ``Exploring protein conformations in vitro and in cell with {EPR}
  distance measurements,'' \emph{Curr. Opin. Struct. Biol.}, vol.~75, p.
  102398, 2022.

\bibitem{xray_a}
P.~V. Afonine, A.~Albert, K.~Diederichs, J.~A. Hermoso, E.~Krissinel, J.~A.
  M\'arquez, S.~Panjikar, M.~Sol\`a, A.~Thorn, and I.~Us\'on, ``Macromolecular
  crystallography,'' \emph{Nat. Rev. Methods Primers}, vol.~5, p.~64, 2025.

\bibitem{xray_b}
I.~Rathore, V.~Mishra, and P.~Bhaumik, ``Advancements in macromolecular
  crystallography: from past to present,'' \emph{Emerg. Top. Life Sci.},
  vol.~5, p. 127, 2021.

\bibitem{cryoem_a}
E.~Nogales and J.~Mahamid, ``Bridging structural and cell biology with
  cryo-electron microscopy,'' \emph{Nature}, vol. 628, p.~47, 2024.

\bibitem{cryoem_b}
M.~Guaita, S.~C. Watters, and S.~Loerch, ``Recent advances and current trends
  in cryo-electron microscopy,'' \emph{Curr. Opin. Struct. Biol.}, vol.~77, p.
  102484, 2022.

\bibitem{deer_benchmark2021}
O.~Schiemann, C.~A. Heubach, D.~Abdullin, K.~Ackermann, M.~Azarkh, E.~G.
  Bagryanskaya, M.~Drescher, B.~Endeward, J.~H. Freed, L.~Galazzo, D.~Goldfarb,
  T.~Hett, L.~Esteban~Hofer, L.~F\'abregas Ib\'a\~nez, E.~J. Hustedt,
  S.~Kucher, I.~Kuprov, J.~E. Lovett, A.~Meyer, S.~Ruthstein, S.~Saxena,
  S.~Stoll, C.~R. Timmel, M.~Di~Valentin, H.~S. Mchaourab, T.~F. Prisner, B.~E.
  Bode, E.~Bordignon, M.~Bennati, and G.~Jeschke, ``Benchmark test and
  guidelines for {DEER}/{PELDOR} experiments on nitroxide-labeled
  biomolecules,'' \emph{J. Am. Chem. Soc.}, vol. 143, no.~43, pp.
  17\,875--17\,890, 2021.

\bibitem{jumper2021}
J.~Jumper, R.~Evans, A.~Pritzel, T.~Green, M.~Figurnov, O.~Ronneberger,
  K.~Tunyasuvunakool, R.~Bates, A.~{\v Z}{\'i}dek, A.~Potapenko \emph{et~al.},
  ``Highly accurate protein structure prediction with {AlphaFold},''
  \emph{Nature}, vol. 596, no. 7873, pp. 583--589, 2021.

\bibitem{polyhach2012}
Y.~Polyhach, E.~Bordignon, R.~Tschaggelar, S.~Gandra, A.~Godt, and G.~Jeschke,
  ``High sensitivity and versatility of the {DEER} experiment on nitroxide
  radical pairs at {Q}-band frequencies,'' \emph{Phys. Chem. Chem. Phys.},
  vol.~14, no.~30, pp. 10\,762--10\,773, 2012.

\bibitem{schops2015}
P.~Sch{\"o}ps, P.~E. Spindler, A.~Marko, and T.~F. Prisner, ``Broadband spin
  echoes and broadband {SIFTER} in {EPR},'' \emph{J. Magn. Reson.}, vol. 250,
  pp. 55--62, 2015.

\bibitem{isaev2023}
N.~P. Isaev, A.~R. Melnikov, K.~A. Lomanovich, M.~V. Dugin, M.~Y. Ivanov, D.~N.
  Polovyanenko, S.~L. Veber, M.~K. Bowman, and E.~G. Bagryanskaya, ``A
  broadband pulse epr spectrometer for high-throughput measurements in the
  {X}-band,'' \emph{J. Magn. Reson. Open}, vol. 14--15, p. 100092, 2023.

\bibitem{kaufmann2013}
T.~Kaufmann, T.~J. Keller, J.~M. Franck, R.~P. Barnes, S.~J. Glaser, J.~M.
  Martinis, and S.~Han, ``{DAC}-board based {X}-band {EPR} spectrometer with
  arbitrary waveform control,'' \emph{J. Magn. Reson.}, vol. 235, pp. 95--108,
  2013.

\bibitem{doll2019}
A.~Doll, ``Pulsed and continuous-wave magnetic resonance spectroscopy using a
  low-cost software-defined radio,'' \emph{AIP Adv.}, vol.~9, no.~11, p.
  115110, 2019.

\bibitem{hill2023}
M.~V. Hanabe~Subramanya, J.~Marbey, K.~Kundu, J.~E. McKay, and S.~Hill,
  ``Broadband {Fourier}-transform-detected {EPR} at {W}-band,'' \emph{Appl.
  Magn. Reson.}, vol.~54, no.~1, pp. 165--181, 2023.

\bibitem{doll2014}
A.~Doll and G.~Jeschke, ``Fourier-transform electron spin resonance with
  bandwidth-compensated chirp pulses,'' \emph{J. Magn. Reson.}, vol. 246, pp.
  18--26, 2014.

\bibitem{doll2017}
A.~Doll and G.~Jeschke, ``Wideband frequency-swept excitation in pulsed {EPR}
  spectroscopy,'' \emph{J. Magn. Reson.}, vol. 280, pp. 46--62, 2017.

\bibitem{jeschke_mlgr}
R.~Tschaggelar, F.~D. Breitgoff, O.~Oberh\"ansli, M.~Qi, A.~Godt, and
  G.~Jeschke, ``High-bandwidth {Q}-band {EPR} resonators,'' \emph{Appl. Magn.
  Reson.}, vol.~48, p. 1273, 2017.

\bibitem{atomize}
A.~Melnikov, A.~Vedkal, A.~Ishchenko, and S.~Veber, ``Atomize: A modular
  software for control and automation of scientific and industrial
  instruments,'' \emph{Journal of Open Research Software}, vol.~13, no.~1,
  p.~26, 2025.

\bibitem{atomize_docs}
A.~Melnikov, ``Atomize documentation,''
  \url{https://anatoly1010.github.io/atomize_docs/}, accessed: 2026-06-29.

\bibitem{hyde1998}
J.~S. Hyde, H.~S. Mchaourab, T.~G. Camenisch, J.~J. Ratke, R.~W. Cox, and
  W.~Froncisz, ``Electron paramagnetic resonance detection by time-locked
  subsampling,'' \emph{Rev. Sci. Instrum.}, vol.~69, no.~7, pp. 2622--2628,
  1998.

\bibitem{stoll2008}
S.~Stoll and B.~Kasumaj, ``Phase cycling in electron spin echo envelope
  modulation,'' \emph{Appl. Magn. Reson.}, vol.~35, no.~1, pp. 15--32, 2008.

\bibitem{numpy}
C.~R. Harris, K.~J. Millman, S.~J. van~der Walt \emph{et~al.}, ``Array
  programming with {NumPy},'' \emph{Nature}, vol. 585, pp. 357--362, 2020.

\bibitem{borodulina2025}
A.~V. Borodulina, A.~R. Melnikov, A.~A. Samsonenko, O.~Y. Rogozhnikova, V.~M.
  Tormyshev, M.~V. Fedin, and S.~L. Veber, ``Monitoring {$S=0 \leftrightarrow
  S=2$} spin-state switching in {Fe(II)} complex using {FT} {EPR} and trityl
  radical as local magnetic field sensor,'' \emph{Appl. Magn. Reson.}, vol.~56,
  no.~11, pp. 1535--1547, 2025.

\bibitem{bowman2021}
M.~K. Bowman and A.~G. Maryasov, ``The direct dimension in pulse {EPR},''
  \emph{Appl. Magn. Reson.}, vol.~52, no.~8, pp. 1041--1062, 2021.

\bibitem{matveeva2021}
A.~G. Matveeva, V.~N. Syryamina, V.~M. Nekrasov, and M.~K. Bowman,
  ``Non-uniform sampling in pulse dipolar spectroscopy by {EPR}: the
  redistribution of noise and the optimization of data acquisition,''
  \emph{Phys. Chem. Chem. Phys.}, vol.~23, pp. 10\,335--10\,346, 2021.

\bibitem{nekrasov2025}
V.~M. Nekrasov, A.~G. Matveeva, V.~N. Syryamina, S.~A. Agarkin, and M.~K.
  Bowman, ``Analytically optimized noise redistribution in pulsed dipolar {EPR}
  spectroscopy,'' \emph{Phys. Chem. Chem. Phys.}, vol.~27, pp. 3272--3281,
  2025.

\end{thebibliography}

\end{document}